\documentclass{aa}
\usepackage{graphics}
\usepackage{epsfig}
\newcommand{\ltsima} {$\; \buildrel < \over \sim \;$}
\newcommand{\gtsima} {$\; \buildrel > \over \sim \;$}
\newcommand{\lta} {\lower.5ex\hbox{\ltsima}}
\newcommand{\gta} {\lower.5ex\hbox{\gtsima}}
\begin{document}
%\setcounter{table}{0}
%\thesaurus{03}
\title{The HST survey of the B2 sample of radio galaxies:\\
detection of two optical jets
\thanks{Based on observations with the NASA/ESA Hubble Space Telescope, obtained at the Space Telescope Science Institute, 
which is operated by AURA, Inc., under NASA contract NAS 5-26555 and by STScI grant GO-3594.01-91A}}
\author{P. Parma\inst{1} 
\and H.R. de Ruiter\inst{2,1} 
\and A. Capetti\inst{3}
\and R. Fanti\inst{4,2} 
\and R. Morganti\inst{5} 
\and M. Bondi\inst{1}
\and R.A. Laing\inst{6,7} 
\and J.R. Canvin\inst{7}}
\institute{Istituto di Radioastronomia, Via Gobetti 101, I-40129, Bologna, Italy
\and Osservatorio Astronomico di Bologna, Via Ranzani, 1, I-40127, Bologna, Italy
\and Osservatorio Astronomico di Torino, Strada Osservatorio 25,  I-10025, Pino Torinese, Italy
\and Istituto di Fisica, Universit\`a degli Studi di Bologna, I-40126, Bologna, Italy
\and Netherlands Foundation for Research in Astronomy, Postbus 2, NL-7990AA, Dwingeloo, The Netherlands
\and Space Science and Technology Department, CLRC, Rutherford Appleton Laboratory, Chilton, Didcot, Oxfordshire OX11 0QX, UK
\and University of Oxford, Department of Astrophysics, Denys Wilkinson Building, Keble Road, Oxford OX1 3RH, UK}
\offprints{P. Parma}
\titlerunning{Optical jets in B2 radio galaxies}
\authorrunning{Parma et al.}
\date{Received 5 August 2002; Accepted 8 October 2002}
\maketitle
\begin{abstract} 
We present HST observations of previously undetected optical jets in the
low-luminosity radio galaxies B2\,0755+37 and B2\,1553+24.  We show that
there is accurate spatial coincidence between optical and radio emission,
implying that the former is likely to be synchrotron radiation.  The physical
properties of the jets are similar to those known previously: their
radio-optical spectral indices are $\approx$0.7 and in B2\,0755+37, the
spectrum steepens between optical and X-ray wavelengths.  Our results support
the hypothesis that optical emission is detectable from jets orientated
within $\approx$20$^\circ$ of the line of sight for the B2 sample.  
\end{abstract}
\section{Introduction}\label{sec:intro} 
Jets are commonly observed in radio galaxies, especially in those with an FR\, I structure (Fanaroff \& Riley \cite{FR74}),
where they are detected in more than half of the objects.  The jets are made visible, at least at radio wavelengths, by
synchrotron emission from relativistic particles moving in magnetic fields. Their physics (e.g. the balance between
particle and magnetic field energies and the particle acceleration processes) is still poorly understood, however. While
hundreds of kpc-scale radio jets are known, only a few have been detected at optical wavelengths. HST has made a great
impact on these studies, not only by having almost doubled the number of known optical jets (Sparks et al.\
\cite{sparks94}; Scarpa \& Urry \cite{scarpa02}), but also because its high spatial resolution, comparable to that of the VLA
at radio wavelengths, makes a direct comparison possible.  Before the present study, 14 optical jets were known. They
are present in all kinds of radio source, but show common properties: they originate from bright unresolved optical nuclei (the one exception is 3C\,15; Martel et al.\ \cite{martel98}); they are smaller and narrower in appearance than their radio
counterparts and no counter-jets are observed. The optically detected jets are longer than expected from their synchrotron
cooling times, and there is no indication of strong steepening of the radio-optical spectral index with increasing distance
from the nucleus (Sparks et al.  \cite{sparks96}; Scarpa et al.  \cite{scarp99a}).  The electrons must therefore be
continually reaccelerated unless the magnetic field is significantly weaker than the equipartition value (Heinz \& 
Begelman \cite{HB}). The advent of {\em Chandra} with its high angular resolution and sensitivity at X-ray wavelengths is 
opening another chapter in the study of jets in FR\,I radio galaxies. Several jets have now been detected by {\em Chandra}
(Hardcastle et al.\ \cite{hardcastle01,hardcastle02}; Worrall et al. \cite{worrall01}; Kraft et al.\ 
\cite{Kraft}; Harris et al. \cite{harris02a,harris02b}; Chiaberge et al.\ \cite{chiaberge02}; Marshall et al. \cite{marshall02}).  
Of these, only M\,87 and 3C66B were known to have prominent optical counterparts (Sparks et al.\ \cite{sparks00} 
and references therein). Are X-ray and optical emission common features of extended jets in all types of radio source? In 
order to address this question it is necessary to carry out a survey of radio jets in order to find their optical and X-ray 
counterparts and to derive their spectral energy distributions.  Here we present the results of a systematic search for 
optical jets in HST images obtained from the snapshot survey of the B2 sample of low-luminosity radio galaxies (Capetti 
et al. \cite{capetti00}).  Of the 57 sources observed with HST, the majority have FR\,I structures and 30 have one- or two-
sided radio jets. We discuss two previously unknown optical jets discovered in the HST images.Throughout the paper we 
take $H_{0} = 100$ km\, s$^{-1}$Mpc$^{-1}$. The  spectral index $\alpha$ is defined in the sense $S(\nu) \propto \nu^{-
\alpha}$.
\section{Observations and Data Analysis}\label{sec:thesample}
Observations and images of the B2 radio galaxies obtained in the course of our HST program were presented in detail in 
Capetti et al.\ (\cite{capetti00}).  Of the $\sim$100 B2 radio galaxies, images were obtained for 41 as part of our 
programme and public archive data were available for a further 16. Observations were obtained using the Wide Field and
 Planetary Camera 2 (WFPC2). The pixel size of the Planetary Camera, on which all targets were centred, is 0.0455 arcsec
and the $800\times 800$ pixels cover a field of view of $36\times 36$ arcsec$^2$. Images with exposure times of 300\,s  
were obtained in each of two broad-band filters: F555W and F814W for the new observations; F555W and F702W for the 
archival data.  Although the transmission curves do not exactly match the standard passbands, we will usually refer to 
them as V and I filters.  The data were processed through the standard Post Observation Data Processing System pipeline
(Biretta et al.\ \cite{biretta}).  We used the task `ellipse' in STSDAS/IRAF to obtain isophotal fits to the light distributions in
all of our images. Regions that were disturbed by the presence of dust or foreground stars were masked before the fitting. 
In several cases, dust prevented a meaningful analysis of the isophotes, particularly in the central portion of the galaxy, 
and no fit was produced.  Whenever possible, we left all of the parameters describing the isophotes free to vary. However,
 in regions of low signal-to-noise, or if the isophotal behaviour was particularly complex, we fixed the centre, ellipticity and
 position angle of the isophotes. Results of the full isophotal analysis of the B2 radio-galaxies will be presented 
elsewhere. Detection of an optical jet is relatively straightforward in the residual image, obtained by subtracting an 
elliptical model fit of the galaxy from the original data. All residual images were smoothed to a resolution of $0\farcs2$ 
to improve their signal-to-noise and searched for statistically significant (positive) residuals aligned at the position angle
 of the radio emission. Optical jets were detected in two objects: B2\,0755+37 and B2\,1553+24.  Their properties are 
summarized in Table \ref{tab:sourcedat}.
\begin{table*}\caption[]{Properties of the Radio Galaxies}
\label{tab:sourcedat}
\begin{flushleft}
\begin{tabular}{lccccc}
\hline\noalign{\smallskip}
Name & Structure & redshift & $\log (P_{\rm tot}$/W\, Hz$^{-1}$) & $\log (P_{\rm core}$/W\, Hz$^{-1}$) & Size \\
B2   &           &          &  1.4 GHz                           & 1.4 GHz               & kpc\\
\noalign{\smallskip}
\hline\noalign{\smallskip}
0755+37 & FRI    & 0.0413  & 24.49              &  23.59        & 76  \\
1553+24 & FRI    & 0.0426  & 23.57               & 23.02        & 162 \\
\noalign{\smallskip}
\hline
\end{tabular}
\end{flushleft}
\end{table*}
\section{Description of the two optical jets }\label{sec:description}
\subsection{B2\,0755+37}\label{subs:075537}
\begin{figure}
\epsfig{file=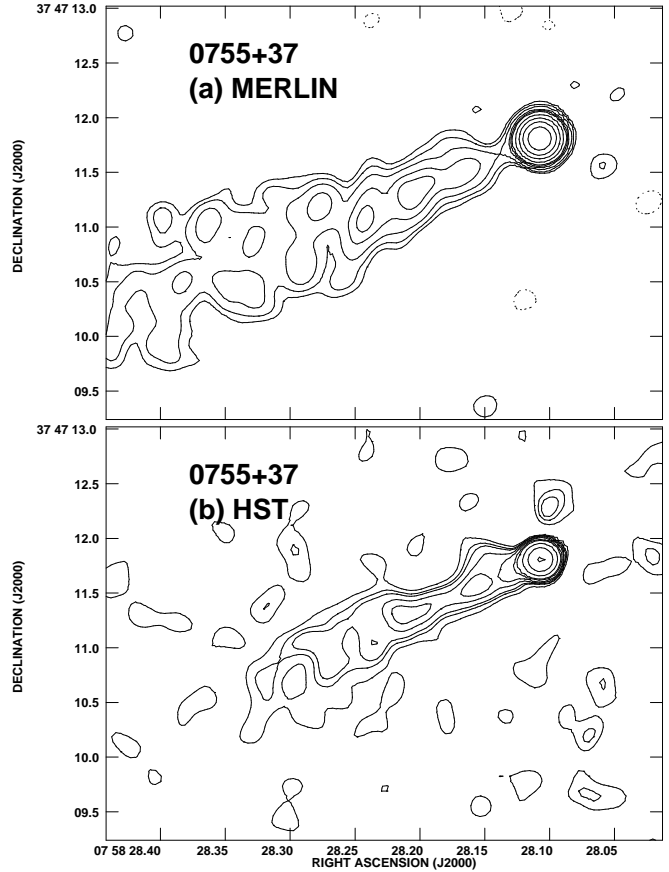,width=9cm,clip}
\caption[]{Radio and optical images of the inner jet of B2\,0755+37 with a resolution of $0\farcs21$ FWHM. 
{\bf a} MERLIN image at 1.658\,GHz (Bondi et al. \cite{bondi00}). The contour levels are $-$1, 1, 2, 4, 8, 16, 32, 64, 128, 256
 $\times$ 0.3 mJy\,/beam area. 
{\bf b} HST image. The relative contour levels are 3, 6, 8, 12, 16, 24, 32, 64, 128, 256.}
\label{fig:b0755}
\end{figure}

\begin{figure}
\epsfig{file=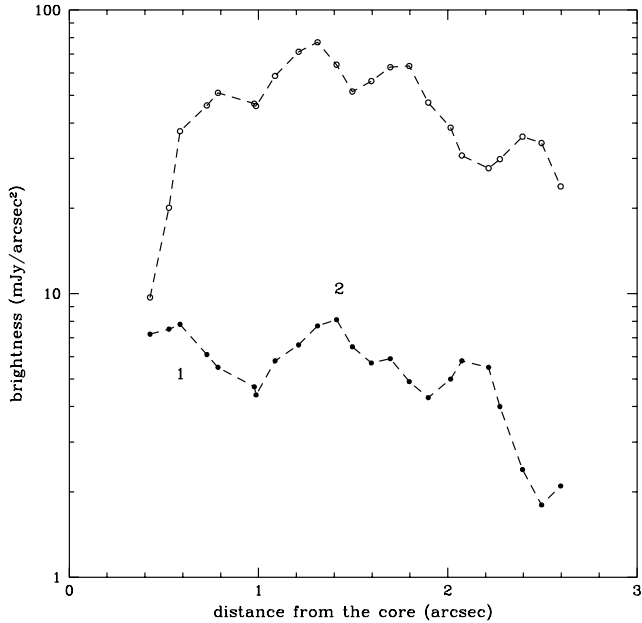,width=9cm,clip}
\caption[]{Radio and optical brightness profiles of the jet of B2~0755+37. The profiles represent the average brightness in 
rectangular areas, with width  10 pixels perpendicular and length 2 pixels parallel to the jet direction. The lower profile
 represents the optical data (scaled upwards by a factor $10^3$), where the numbers refer to knots 1 and 2. The upper 
profile represents the radio jet.}
\label{fig:ridge_line}
\end{figure}

\begin{figure}
\epsfig{file=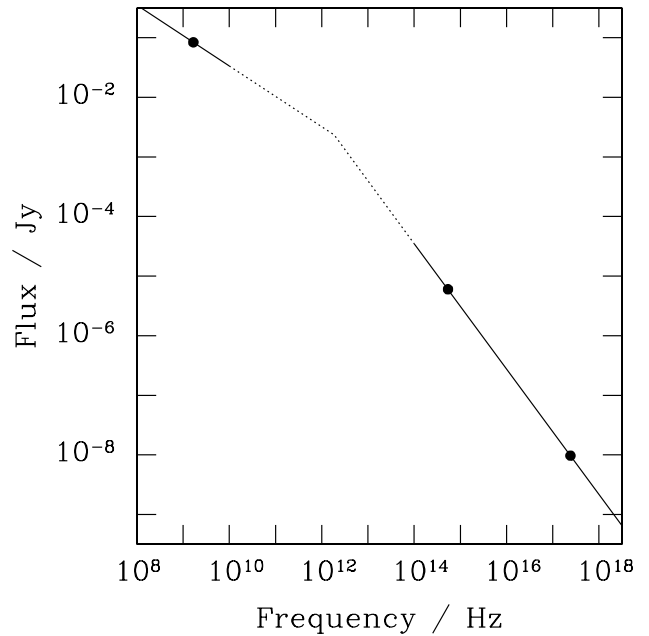,width=9cm,clip}
\caption[]{The broad-band spectrum of a 4\,arcsec $\times$ 1.5\,arcsec region at the base of the jet in B2\,0755+37.  The 
data-points are at 1658\,MHz (Bondi et al.\ \cite{bondi00}), 0.555\,$\mu$m (present paper) and 1\,keV (Worrall et al.\ 
\cite{worrall01}).  Indicative power-law spectra   with $\alpha = 0.51$ and 1.05 have been drawn through the radio
 and optical--X ray points, respectively.}
\label{fig:0755spectrum}
\end{figure}

This well studied nearby object (Bondi et al.\ \cite{bondi00} and references therein) is an FR\,I source with two large 
symmetrical lobes and very asymmetric jets, the brighter of which is better collimated.  The bright base of the main jet 
has been studied with high spatial resolution on scales from few pc to several kpc using the VLA, MERLIN and VLBI and 
appears to be very well collimated. At the highest MERLIN resolution it is barely resolved across its width.  At a distance
 of $\sim 0.5 $ arcsec from the core the jet flares abruptly (Fig.~\ref{fig:b0755}a) and its brightness distribution becomes 
much smoother.
In the optical band, the galaxy has a rather smooth appearance, without any sign of dust obscuration. A bright optical 
nucleus is seen at its centre (Capetti et al. 2002).  In both the V and I residual images, a well defined linear structure, 
coincident with the radio jet, extends for $\approx$3\,arcsec from the nucleus.  The overall appearance of the optical jet 
(Fig. \ref{fig:b0755}b), is similar that of the radio jet.  The bright structure near the nucleus, which is partly blended with the
 central point source, corresponds closely to the base of the radio jet.  In addition there are two bright knots embedded in 
a low brightness structure, which widens further away from the core.
The similarity between the optical and radio features, as can be seen in Fig.~\ref{fig:ridge_line}, suggests that the
 observed optical emission is synchrotron radiation but, as in 3C\,66B (Macchetto et al.\ \cite{macchetto91}), the radio 
emission extends much further from the nucleus. 
This is mainly due to the limited sensitivity of the optical image: the brightness of the optical jet fades like the radio jet, 
but reaches the background much earlier.
The radio and optical fluxes were determined over identical areas of the 1658-MHz MERLIN and HST images, respectively,
 after convolution to the same resolution. We then derived the radio-to-optical spectral index for the two knots and for the 
entire jet (Table \ref{tab:jetdat}). Note that the uncertainties in the radio to optical, or optical to X, spectral indices are small
 because of the long frequency baselines: (one $\sigma$) errors of the order of 3 \% in flux density result in 3$\sigma$ 
errors in the slope of 0.01. Errors of this magnitude apply to B2~0755+37, while our worst case (a $\sim 20$ \% uncertainty
 in the optical flux of the jet of B2~1553+24) gives a 3$\sigma$ error $\Delta\alpha \sim 0.04$ (see 
Table~\ref{tab:jetdat}). Based on these data the difference in spectral index between knots 1 and 2 is highly 
significant. Worrall et al. (\cite{worrall01}) show asymmetric X-ray emission coincident with the base of the radio jet. Using
 the X-ray flux of the jet quoted by Worrall et al. ( \cite{worrall01}), together with radio and optical fluxes derived for 
the same region (a box of 4 $\times$ 1.5\,arcsec$^2$ with the long axis along the jet), we derived the spectrum shown in 
Fig.~\ref{fig:0755spectrum}.  The spectral index between 1.7 and 4.9\,GHz is $\alpha_{\rm R} = 0.51$ (Bondi et al. 
\cite{bondi00}). The spectrum steepens with increasing frequency, with $\alpha_{\rm RO} = 0.74\pm 0.01$ and 
$\alpha_{\rm OX} = 1.05\pm 0.01$.
\subsection{B2\,1553+24}\label{subs:155324}
\begin{figure}\epsfig{file=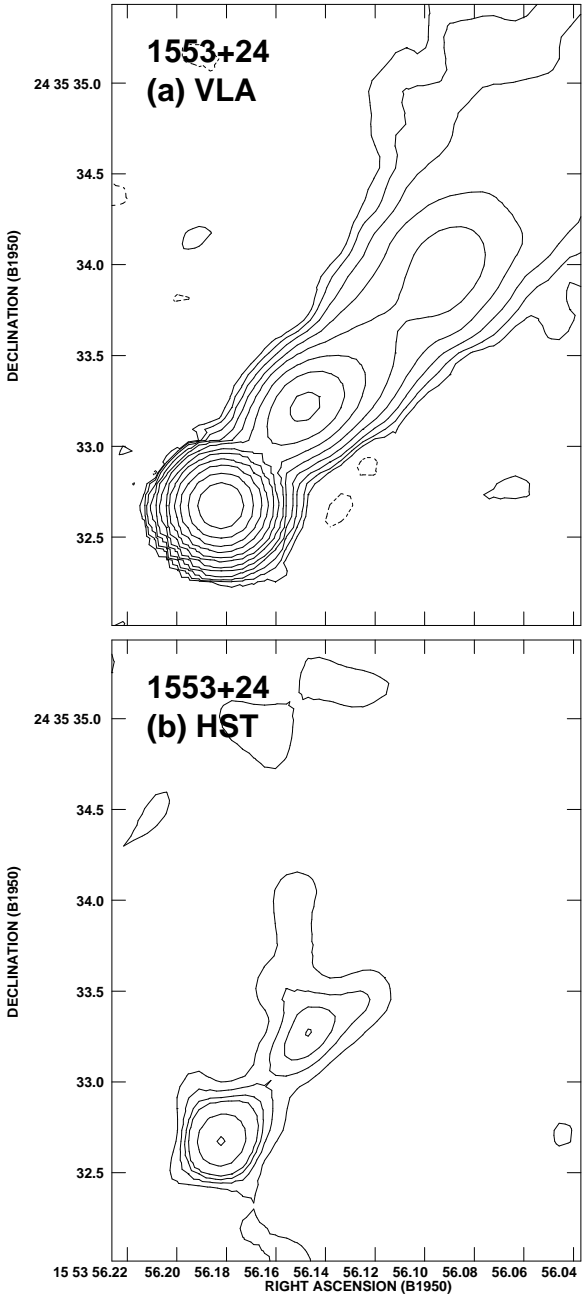,width=8cm,clip=}
\caption[]{Images of the inner jet of B2\,1553+24 at a resolution of $0\farcs25$ FWHM.  {\bf a} VLA image at 8.4\,GHz
 (Canvin \& Laing, in preparation). The contour levels are $-$3, 3, 6, 12, 24, 48, 100, 200, 400, 800, 1600, 3200, 6400 
$\times$ 5\,$\mu$Jy/beam area.  {\bf b} HST image. The relative contour levels are 3, 6, 9, 12, 24, 48}
\label{fig:b1553}
\end{figure}
This FR\,I radio source was originally classified as having ``naked jets'' without lobe emission (Parma et al.\ 
\cite{parma87}).  Very faint lobes have subsequently been found (Stocke \& Burns \cite{stocke87}), but the prominence 
of the jet structure compared with the diffuse extended components is unusually high. 
The source has been observed at the VLA at 1.4 and 4.8\,GHz in several configurations (Morganti et al.\ \cite{morg87}).  At 
low resolution, the two jets are quite similar, but higher-resolution images reveal a pronounced asymmetry close to the 
nucleus.  A portion of a recent deep VLA image combining data from the A, B and C configurations at 8.4\,GHz (Canvin \&
 Laing, in preparation) is shown in Fig. \ref{fig:b1553}a. The resolution is 0.25\,arcsec. At the base of the main jet, 
$\approx$0.85\,arcsec from the nucleus, there is a single bright knot. After a secondary knot at $\approx$1.8\,arcsec, the 
jet expands more rapidly.  No counter-jet is seen out to the point at which the jet recollimates. In the HST images, the host
 galaxy of B2\,1553+24 has a very smooth morphology, with a bright optical nucleus, just as in B2\,0755+37.  The residual
 images show a one-sided jet (Fig. \ref{fig:b1553}b) whose structure is mainly defined by the presence of an elongated 
knot at 0.85\,arcsec (0.51 kpc) from the nucleus, coincident with the first bright radio knot.  The knot, whose parameters 
are given in Table \ref{tab:jetdat}, is resolved both along and perpendicular to the jet axis. Its radio-to-optical spectral 
index $\alpha_{\rm RO} = 0.79$.  Again, the precise positional coincidence supports a synchrotron interpretation for the
 optical emission.
\section{Discussion and Conclusions}\label{sec:discussion}
\subsection{Summary of observed and intrinsic properties}
\label{props}
 In Table \ref{tab:jetdat} we report the optical flux, the distance of each knot from the central point source and the jet and 
knot sizes at these positions. The knot sizes are averages of fitted Gaussian FWHM along and transverse to the axis.  For
 the jet in B2\,0755+37, which is resolved orthogonal to its axis, the reported size refers to the largest value measured on
 the contour map at a level of three times the background rms.
\begin{table*}
\caption[]{Jet data}
\label{tab:jetdat}
\begin{flushleft}
\begin{tabular}{llcrrrrrrr}
\hline\noalign{\smallskip}
Name &Structure& Band & $S_{\rm opt}$ & Distance & Size & $S_{\rm rad}$ &$\alpha_{\rm RO}$& Volume  & B \\
B2   &           &      & $\mu$Jy & arcsec   & arcsec & mJy      & $\pm 3\sigma$      & kpc$^3$ & $\mu$G \\
\noalign{\smallskip}
\hline\noalign{\smallskip}
0755+37 & jet   & V & 5.50 &      & 2.5$\times$0.6 & 68.1 & $0.74\pm 0.01$& 0.1330 &  94 \\
0755+37 & blob1 & V & 0.80 & 0.59 & 0.25           &  3.9 & $0.67\pm 0.01$& 0.0103 &  86 \\
0755+37 & blob2 & V & 1.00 & 1.41 & 0.28           &  8.6 & $0.72\pm 0.01$& 0.0150 &  98 \\
1553+24 & blob  & V & 0.68 & 0.85 & 0.21           &  4.4 & $0.79\pm 0.04$& 0.0067 & 147 \\
\noalign{\smallskip}
\hline\end{tabular}
\end{flushleft}
\end{table*}
Leaving aside Doppler corrections for the moment, and assuming equipartition, we can estimate the internal magnetic 
field. The jet and the knots are assumed to be cylindrical and spherical respectively.  In Table \ref{tab:jetdat} we give the 
volumes V of the individual knots and, for B2\,0755+37, of the entire optical jet.  We assume that the synchrotron 
emission can be described by power laws with spectral indices as given in Table \ref{tab:jetdat}, extending from $\nu = 
10^{7}$\,Hz to $\nu = 10^{15}$\,Hz.  With these assumptions the equipartition fields $B_{\rm eq}$ range from $\approx$100 
to $\approx$150\,$\mu$Gauss, the higher values being found in the more compact knots. The spectrum of B2\,0755+37 
probably breaks at lower frequency  ($\nu = 10^{15}$) but a spectral index   between 1.7 and 4.9\,GHz of $\alpha_{\rm R} = 
0.51$ makes the value of magnetic field insensitive to the limits. Modeling of these jets suggests that both are orientated fairly close to the line of sight and have relativistic velocities. For B2\,0755+37, Bondi et al. (\cite{bondi00}) suggest that
 $\theta \approx$ 27$^\circ$ and $\beta_{\rm 0.5 kpc} \approx 0.9$, while for B2\,1553+24 Canvin and Laing (in 
preparation) infer $\theta$ \lta 15$^\circ$ and $\beta_{\rm 0.5 kpc} \approx 0.8$. This implies that the observed 
luminosities are probably amplified by Doppler beaming, with Doppler factors $\delta \approx 2$. Correcting the observed 
luminosities for the beaming amplification would decrease $B_{\rm eq}$ by roughly a factor of $\delta$. These values are fully
 consistent with those computed for other optical jets (Scarpa and Urry \cite{scarpa02}).
\subsection{Doppler beaming and the detectability of optical jets}
Of the 57 B2 radio sources observed with HST, two were found to have an optical jet above the detection limit ($\sim 2.5 
\mu$Jy/arcsec$^2$ for the majority of the images). However, this fraction goes up to 2/30 (7\%) if we consider only sources
 with radio jets, and to 2/15 (14\%), if we further restrict ourselves to sources with radio core power higher than the median 
value at given extended luminosity (normalized core power $P_{\rm CN}>1$; de Ruiter et al.\ \cite{deruiter90}).  B2\,0755+37 
and B2\,1553+24 also have the highest values of $P_{\rm CN}$ and jet/counter-jet ratio of the 17 sources in common with the 
sample analysed by Laing et al.\ (\cite{laing99}). This is of course consistent with the hypothesis that optical jet emission 
is only detected if the jets are close to end-on: the normalized core power, jet surface brightness and jet/counter-jet ratio 
are all expected to increase with decreasing angle to the line of sight as a result of Doppler beaming (Laing et al. 
\cite{laing99}). If the observed sample is orientated randomly, then the detection rate for sources with radio jets (2/30)
implies that optical jets are seen if $\theta$ \lta 20$^\circ$ (with a statistical uncertainty of at least 10$^\circ$), roughly consistent with the angles derived from models of individual sources (Sect. \ref{props}).  We would expect all sources with
 detected optical emission to have $P_{\rm CN} > 1$, as this corresponds to $\theta <$60$^\circ$ if there is no dispersion in
 the intrinsic core/extended flux ratio. This is indeed the case. The spectra of the detected optical jets are consistent with
 radio-to-optical spectral indices $\alpha_{\rm RO} \approx 0.7$ with relatively little deviation from power laws (implying 
that the Doppler shift of the spectrum due to relativistic motion will be a second-order effect).  If all of the radio jets in the 
B2 sources have the same spectra, then we can predict their optical fluxes given the (known) radio fluxes.  We expect no 
further optical detections to our flux limit. The source that comes closest is B2\,1521+28, for which the expected optical 
brightness is around 1 $\mu$Jy/arcsec$^2$, implying that detection might have been possible for a slightly flatter spectral
index. These results therefore suggest that it is relatively easy to predict the likelihood of detection of an optical jet once
 we know the surface brightness of the radio jet. Sparks et al.\ (\cite{sparks95}) claim that sources with detected optical jets
 tend to be significantly smaller than those without (the median sizes are 70--80 kpc and 200 kpc respectively), as 
expected if the optical jet sources are close to the line of sight.  Our two new detections do not provide support for this
 idea, since their sizes are 76 kpc (B2\,0755+37) and 162 kpc (B2\,1553+24), but the dispersion in intrinsic linear size is 
sufficiently large that the result is inconclusive.  
\subsection{The X-ray emission mechanism in B2\,0755+37}
Our results (Fig.~\ref{fig:0755spectrum}) show that the broad-band spectrum of the jet in B2\,0755+37 is very similar to that
 for comparable structures in other FR\,I radio galaxies such as 3C\,66B (Hardcastle et al. \cite{hardcastle01}), M\,87 
(Marshall et al.\ \cite{marshall02}), 3C\,31 (Hardcastle et al.\ \cite{hardcastle02}) and Cen\,A (Kraft et al.\ \cite{Kraft}). In all 
cases, the spectra steepen from $\alpha \approx$ 0.5 -- 0.6 at radio frequencies to $\alpha \approx$ 1.0 -- 1.3 in the X-ray band.  The shape of the spectrum is consistent with Worrall et al.( \cite{worrall01})'s conclusion that the X-ray emission 
mechanism is synchrotron radiation. As they point out, synchrotron self-Compton emission for an equipartition magnetic
 field is expected to be three orders of magnitude weaker than is observed. Inverse Compton scattering of the microwave 
background is also ruled out for the estimated angle to the line of sight ($\theta \approx$ 27$^\circ$), as extremely large
 beaming factors would be required.
\subsection{Conclusions}
We have presented HST observations of two previously undetected optical jets in low-luminosity FR\,I radio sources, 
B2\,0755+37 and B2\,1553+24.  In both objects, the optical jet emission is one sided and dominated by bright knots. 
In B2\,0755+37 we also see more extended low-brightness emission.
The values of the radio-optical spectral index and $B_{\rm eq}$ are similar to those found in other optical jets. In the jet of B2\,0755+37 there is a slight indication of a steepening of the spectrum with increasing distance from the nucleus, but 
deeper optical images are needed to investigate this effect in detail.  The spectrum of the jet also steepens between 
optical and X-raywavelengths, strengthening the evidence for synchrotron X-ray emission. Both sources show bright optical cores and there is no evidence for circumnuclear dust.  Our results are consistent with the idea that optical jets are
 currently detectable in sources with $\theta$ \lta 20$^\circ$, where emission from the approaching jet is enhanced by 
Doppler beaming.
\begin{acknowledgements}This research has made use of the NASA/IPAC Extragalactic Database (NED) which is operated 
by the Jet Propulsion Laboratory, California Institute of Technology, under contract with the National Aeronautics and 
Space Administration.  The National Radio Astronomy Observatory is a facility of the National Science Foundation 
operated under cooperative agreement by Associated Universities, Inc.
\end{acknowledgements}

\end{document}